\documentstyle{mn}
\newcommand{\reference}{\bibitem}
\def\beq{\begin{equation}}
\def\eeq{\end{equation}}
\def\bey{\begin{eqnarray}}
\def\eey{\end{eqnarray}}
\def\beqarray{\begin{eqnarray}}
\def\eeqarray{\end{eqnarray}}

\def\v200{V_{200}}
\def\rv{r_v}

\def\rr{\left({r \over r_v}\right)}

\def\Mtot{{M_{\rm tot}}}
\def\mtot{{m_{\rm tot}}}
\def\jtot{{j_{\rm tot}}}
\def\rd{{r_{\rm es}}}

\def\arot{{\alpha_{\rm rot}}}

\def\Mtot{{M_{\rm tot}}}
\def\rmax{r_{\rm max}}
\input epsf
\title[]	
{The Origin of the Hubble Sequence for Spiral Galaxies}

\author[]	
{Shude Mao, H.J. Mo
\thanks {E-mail: (smao, hom)@mpa-garching.mpg.de} \\
      Max-Planck-Institut f\"ur Astrophysik
      Karl-Schwarzschild-Strasse 1, 85748 Garching, Germany}
\date{Accepted ........
      Received .......;
      in original form .......}
\pubyear{1998}

\begin{document}
\maketitle
\begin{abstract}
 We suggest that the bulge-to-disc ratios of spiral galaxies 
are primarily determined by the angular momenta of 
their host haloes predicted in current hierarchical clustering models 
for structure formation. Gas with low specific angular momentum
becomes self-gravitating and presumably forms stars before it can
settle into a rotationally supported disc. We assume this part of
the gas in a dark halo to form a bulge, while the rest 
is assumed to settle into a rotationally supported disc.
With these assumptions the predicted bulge-to-disc ratios in 
mass and in size, and other correlations between the bulge and 
disc components can match current observational results. 
This model predicts the existence of a population of 
low-surface-brightness galaxies which are bulgeless.
The model also predicts that the bulge component
has many properties in common with the disc component, 
because both form through similar processes. 
In particular, many bulges should be supported 
(at least partially) by rotation. 
\end{abstract}

\section {Introduction}

 One fundamental observation on galaxy structure is that
spiral galaxies typically have two components, a disc and
a bulge. The bulge-to-disc ratio (in luminosity)
is considered to be an important 
factor in determining the morphological types in the Hubble 
sequence. An important goal of galaxy formation  
is therefore to understand how the bulge and disc
components form and what determines the bulge-to-disc 
ratio. The formation of galactic discs 
has been studied quite extensively in the literature
(e.g. Fall \& Efstathiou 1980; Dalcanton, Summers \& Spergel
1997; Mo, Mao \& White 1998, hereafter MMW; Jimenez et al 1998). 
As discussed in MMW, disc formation can 
be understood as a result of
gas settling into dark haloes predicted
by current hierarchical clustering models of structure formation.
In contrast, the formation of galactic bulges is still ill
understood (see Wyse, Gilmore \& Franx 1997 for a review).
Three classes of scenarios have been suggested.   
The first class assumes that bulges form from early 
collapse of a protogalaxy (Eggen et al 1962).
In this scenario, the size and mass of the bulge component 
of a galaxy depends on the efficiency of the first burst of
star formation, while the properties of discs   
are determined by later infall of high angular momentum 
material.  The second class assumes that bulges form from 
mergers of disc galaxies (Toomre \& Toomre 1972;
Kauffmann, White \& Guiderdoni, 1993; Kauffmann 1996).
In this scenario, the bulge-to-disc ratio 
is determined by the amount of gas locked into stars at the time 
of last major merger. Some quantitative predictions
have been given in Kauffmann (1996). The third class 
assumes that bulges form as a result of the secular evolutions of 
galactic discs. In this scenario, the (inner) thin disc is transformed
into a bulge by vertical heating due to, e.g. 
the bar instability (Combes et al 1990; Norman, Sellwood, \& Hasan
1996, and references therein). 
%This scenario has gain received some recent observational support since the
%disc and bulge properties are correlated
%(Courteau, de Jong \& Broeils 1996 and references therein).
Unfortunately, none of the models predicts the relative 
bulge-to-disc size distribution, and with the exception of 
the merger model (Kauffmann 1996),
the theoretical models have not been carried out to the point
of quantitative predictions about the relative proportion of 
bulge and disc and other observed correlations between the 
bulge and disc components.

 In this paper, we suggest that the bulge-to-disc ratios 
of spiral galaxies are primarily determined by the angular momentum of 
dark haloes predicted in current hierarchical clustering models 
for structure formation. This model (described in Section 2) allows
one to put the formation of the Hubble sequence for spiral galaxies 
into cosmological context. It also makes quantitative predictions for 
the properties of the bulge-disc systems (Section 3). 
The implications of our results are discussed in Section 4.

\section {Model}

In the standard hierarchical clustering models of structure 
formation, dissipationless dark matter particles
are clustered into larger and larger clumps (dark haloes)
in the passage of time due to gravitational
instability. Gas associated with such haloes condenses and 
cools, eventually forming the visible galaxies. The
properties of dark matter haloes are well understood since
they are primarily determined by gravitation and can be readily studied
with numerical simulations. We shall use these properties in our modelling.
On the other hand, the behaviour of the gas 
component and the associated star formation are poorly 
understood. We attempt to model this part by making simple but 
plausible assumptions.

We model dark haloes as spherical objects with
a radial density profile given by
\beq \label{rho}
\rho(r) \propto {1 \over (r/\rv)^{\alpha} (1/c+r/\rv)^2},
\eeq
where $c$ is a parameter that describes the concentration of the halo and
$\rv$ is the virial radius of the halo.
%and $\delta_c$ is a characteristic overdensity that is of no concern here.
Navarro, Frenk \& White (1996, 1997, hereafter
NFW) found through numerical simulations
that $\alpha \approx 1$ and $c \sim 10$ (for galactic-sized 
haloes). We take these as our canonical values, but later vary 
the value of $\alpha$ to see the change in our results. 
Dark haloes acquire angular momentum due to tidal forces from 
nearby structures. The angular momentum of a dark matter halo is 
usually parameterised by a dimensionless quantity 
\beq \label{lambda}
\lambda = {J E^{1/2} \over G M^{5/2}},
\eeq
where $J, E$ and $M$ are the total angular momentum, energy and mass
of the halo, respectively. The distribution of $\lambda$ for dark
haloes in $N$-body simulations are well fitted by a log-normal 
function with mean ${\overline \lambda}=0.05$ and dispersion
$\sigma_\lambda=0.5$ (see equation [15] in MMW; 
Warren et al 1992; Cole \& Lacey 1996; Lemson \& Kauffmann
1998). This distribution depends only weakly 
on cosmology and on the mass and environment of haloes (Lemson 
\& Kauffmann 1998). The variation of specific angular momentum
with radius in a dark halo follows approximately a power-law
\beq \label{jr}
{\cal J}(r) \propto \rr^\beta,
\eeq
where $\beta \approx 1.3$ (Tormen, Mo \& Mao 1998). Notice that the
specific angular momentum is lower at smaller radius because 
material with lower ${\cal J}$ naturally falls into the halo centre. 
Any angular momentum transfer during the gas collapse phase 
probably transports angular momentum outwards and makes the 
effective value of $\beta$ larger. Equations (\ref{rho}-\ref{jr}) 
complete the description for the dark matter
component in our model. 

To model the gas component, we assume that the baryons in a dark 
halo initially have the same mass profile and the same specific
angular momentum profile as the dark matter component.
The total baryon mass that settles into a galaxy is assumed to be 
a constant fraction, $\mtot$, of the halo mass. We take $\mtot=0.05$, as in
MMW. MMW also showed that if this mass settles into a 
rotationally supported exponential
disc with a flat rotation curve,
then the (exponential) scalelength (indicated by subscript es)
is given by
\beq \label{rd}
\rd = {1\over \sqrt{2}} {j_d \over m_d} \lambda \rv,
\eeq
where $m_d\equiv M_d/M$, $j_d\equiv J_d/J$, with
$M_d$ and $J_d$ being the mass and angular momentum 
of the disc. Here we have ignored a factor of order unity that 
depends on halo profile and on disc self-gravity (cf. MMW). 

In order to separate the bulge and disc components, we need to
understand how the gas at different radii in a halo 
settles into the halo centre. We assume that galaxies form
from inside out, so that the scaling in equation (\ref{rd})
can also be applied to the gas within any spheres with  
$r < \rv$. The mass and angular momentum within radius $r$ are
\beq
M(<r)=\int_0^r \rho(r)~ 4\pi r^2 dr,~~~~
J(<r)=\int_0^r {\cal J}(r) ~ dM.
\eeq
The fractional mass and angular momentum within a sphere 
of radius $r$ are therefore
\beq \label{lambda-eff}
j_d(<r) = \mtot {J(<r) \over J}, ~~~~
m_d(<r) = \mtot {M(<r) \over M},
\eeq
where $\mtot$ is (again) the fraction of total halo mass that 
settles into the centre to form the bulge-disc system.
If we assume that the mass in the sphere also settles into a 
rotationally supported exponential profile
conserving angular momentum, then using equation (\ref{rd}) we obtain
\beq \label{rd-sphere}
\rd(<r) = {1\over \sqrt{2}} {j_d(<r) \over m_d(<r)}\lambda \rv. ~~~
\eeq
In practice, we multiply the above scalelength by a factor of
$\arot=1.5$ to take into account the fact that bulges are
only partially rotationally supported 
(e.g. Kormendy \& Illingworth 1982).
Our results, however, are not sensitive to the choice of $\arot$.

Larson (1976), among others,
argued that the bulge component is likely to have formed in
rapid episodes of star formation.
A minimum condition for rapid star formation is
that the gas must become self-gravitating and be able to fragment. 
This means that its density must reach to a level of about 3 times 
the halo density, so that the fragmented clouds are not subject 
to tidal disruption. Similar condition may lead to the
formation of a bar (which then desolves into a bulge
according to the secular-evolution scenario), because
self-gravitating discs may be subject to bar instability 
(e.g. Efstathiou, Lake \& Negroponte 1982).
If the bayron mass within an initial radius 
$r_i$ settles within a final radius $r_f$, the
self-gravitating condition can be written as
\beq \label{gravity}
\mtot M(<r_i) \geq 3 M(<r_f).
\eeq
Since we assume the mass settles into an exponential profile 
which does not have a sharp cutoff radius, we take $r_f$ as the 
effective radius within which half of the mass is enclosed (for an
exponential profile, $r_f=1.67 r_{\rm es}$). Correspondingly,
we take the mass at the left hand side as the baryonic mass enclosed
within the effective radius.
The bulge mass and radius are then determined by the maximum initial 
radius, $\rmax$, which still satisfies the inequality (\ref{gravity}).
Once $\rmax$ is found, the masses and 
scalelengths for the bulge and disc components are given by
\beq 
 M_b = \mtot M(<\rmax),~~
r_b = {1\over \sqrt{2}} \arot {j_d(<\rmax) \over m_d(<\rmax)} \lambda \rv, ~
\eeq
\beq
M_d = \mtot M- M_b,~~
r_d = {1\over \sqrt{2}} {{\jtot}-j_d(<\rmax) 
\over \mtot-m_d(<\rmax)} \lambda \rv,
\eeq
where $\jtot=\mtot$ because we assume that the gas
which forms the bulge-disc system has the same specific
angular momentum as the dark matter halo and has its angular
momentum conserved during the collapse.

\section{Model Predictions}

In our model, there are two parameters, $\alpha$ and $\beta$, 
that describe the runs of mass and specific angular momentum with 
radius. Once $\alpha$ and $\beta$ are given, the predicted
bulge-to-disc scalelength ratio, $r_b/r_d$, and bulge-to-total mass
ratio, $M_b/\Mtot (\Mtot \equiv M_b+M_d)$, depend only
on the spin parameter $\lambda$. In this section, we present our
model predictions and compare them with the recent observations
of de Jong (1996a,b).  De Jong's sample contains 86 nearby 
late-type galaxies that are chosen from the UGC catalogue. 
For each galaxy, various decompositions of the observed 
light into a disc component and a bulge component have been
carried out. It appears that the observed galaxies are
best described by an exponential disc and an exponential 
bulge (see also Courteau, De Jong \& Broeils 1996).
The bulge-to-disc ratios in both scalelength and light are 
given for each galaxy in many bands, which makes it possible for us 
to compare model predictions directly with observations. We use the
I-band data but the results are similar for other bands.

Figure 1 shows the predicted values of $r_b/r_d$ (bottom panel) and
$M_b/\Mtot$ (top panel) as functions of $\lambda$ for
different combinations of $\alpha$ and $\beta$.
These two ratios vary with $\lambda$ in very similar manner. 
The solid lines show the cases with $\beta=1.3$ for three values
of $\alpha$: 0.7, 1 and 1.3. For all combinations of $\alpha$ and
$\beta$, the predicted bulge-to-disc ratio decreases with
increasing $\alpha$. This dependence can be understood as follows.
A larger value of $\alpha$ implies a more concentrated distribution 
of dark matter and therefore requires the gas
to contract more to become self-gravitating.
The two dashed curves show the effect of changing
$\beta$. For a given $\lambda$, the
disc-to-bulge ratio increases with $\beta$, 
because a larger value of $\beta$ implies that
there is more material with low angular momentum 
in the central part to form a bulge. 

\begin{figure}
\epsfysize=9.5cm
\centerline{\epsfbox{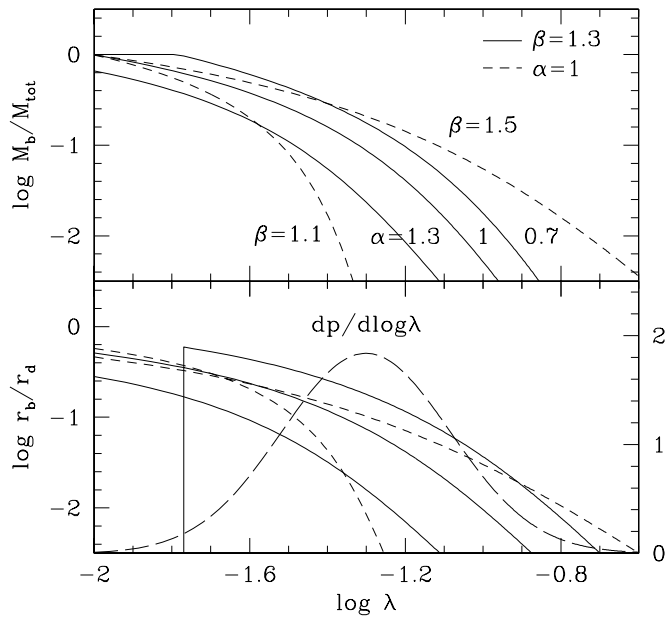}}
\caption{The top panel shows the bulge-to-total mass ratio
as a function of spin
parameter $\lambda$. The bottom panel shows the bulge-to-disc
scalelength ratio as a function of $\lambda$. The solid curves all
have $\beta=1.3$ and their $\alpha$ parameters are labelled besides
the curves. The two dashed lines have $\alpha=1$ and $\beta=1.1, 1.5$
respectively. The ordering of curves in the lower panel is the same as
that in the top panel. The $\lambda$-distribution in haloes is indicated
by the long dashed line in the bottom panel with the scales shown in the
right $y$-axis. The sudden dropoff in the curve for $\alpha=0.7,
\beta=1.3$ indicates that all systems with $\log \lambda \leq -1.78$ form
only bulges; such systems are rare, however.
}
\end{figure}

To have more direct comparisons with the observational results, 
we convolve the predicted bulge-to-disc ratios with the 
$\lambda$-distribution found in numerical simulations. 
This is done by Monte Carlo simulations in which
the value of $\lambda$ is randomly drawn from its probability 
distribution. We fix the values of $\alpha$ and $\beta$ to be 
1 and 1.3 respectively. We also assume that the stellar 
mass-to-light ratio is similar for the disc and bulge
components in a galaxy (see Fig. 1 in de Jong 1996c),
so that the predicted bulge-to-disc
ratio in mass can be compared directly with the observed
bulge-to-disc ratio in luminosity.

The predicted distributions of $r_b/r_d$ and $M_b/\Mtot$ are shown
as the thick solid histograms. Compared with the observational data of
de Jong (1996, shaded histograms), it is clear that the 
median values of the two distributions predicted by our model
are similar to those observed. However, the model predicts 
more systems at both the low and high ends of $r_b/r_d$
and $M_b/\Mtot$. It is not surprising that the predicted 
abundance of systems with large bulge-to-disc ratios is higher
than that given by the observation, because the de Jong sample 
is biased against early-type galaxies which have
larger bulge-to-disc ratios. 
The predicted tails of small bulges are produced mainly by systems with
$\lambda>0.08$. Such systems may form low
surface brightness systems (cf. MMW for a discussion) 
which are probably also missed in the de Jong sample. 
In fact, the low surface brightness galaxies in O'Neil et al 
(1997) sample have remarkably small central bulges.
The discrepancy between the model prediction and 
the observational results may just be due to the selection
bias in the sample.

\begin{figure}
\epsfysize=9.5cm
\centerline{\epsfbox{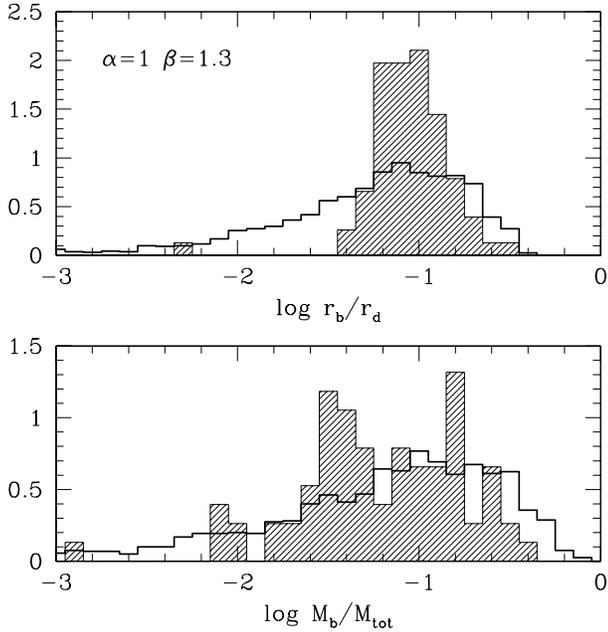}}
\caption{Histograms (thick solid)
for the predicted bulge-to-total mass ratio (lower panel)
and the bulge-to-disc ratio in scalelength (upper panel). The model assumes
$\alpha=1$ and $\beta=1.3$. The shaded
histograms are the data from de Jong (1996a). Note that 
the model predicts more systems at both the small and large 
bulge-to-disc ratio ends (see text).
}
\end{figure}

The galaxies in the de Jong sample also show a well-defined
correlation between the bulge and disc luminosities (cf. Fig. 19
in de Jong 1996b). To see whether our model can reproduce such a trend,
we construct a `galaxy sample' in which each galaxy is given 
a magnitude randomly drawn from the CfA luminosity
function (Marzke et al 1994). Our results remain the same 
for any other luminosity function. 
The top panel of Figure 3 shows the predicted trend for
$\alpha=1$ and $\beta=1.3$. The two solid lines bracket the observed
scatter ($\sim 4$ mag). It is interesting to see that virtually all the
systems with $\lambda<0.08$ (filled dots)
are bracketed within the observed range.
The systems with $\lambda>0.08$ should have smaller 
bulge-to-disc ratios. These systems have systematically
lower surface brightness since their discs are more spread-out 
(MMW). In reality, there should be scatters in the values of
$\alpha$ and $\beta$ for different haloes. Unfortunately
it is unclear how large these scatters are. 
To illustrate how such scatters affect our results, we show 
in the bottom panel of Fig. 3 an example
where $\alpha$ is allowed to vary from 0.7 to 1.3, and $\beta$ from 1.1
to 1.5. The scatter in the bulge-disc luminosity relation 
now becomes bigger; in particular some systems 
with small $\lambda$ can now have very small bulge-to-disc ratios.

\begin{figure}
\epsfysize=9.5cm
\centerline{\epsfbox{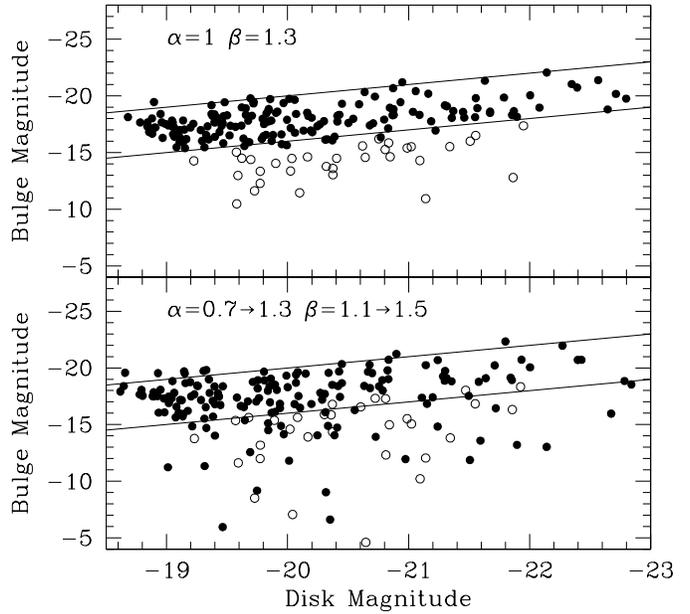}}
\caption{Predicted correlation between
the bulge magnitude vs. disc magnitude.
The top panel is for Monte Carlo simulations with
$\alpha=1, \beta=1.3$ while the bottom one shows the results where
$\alpha$ is allowed to vary uniformly between  0.7 to 1.3 and $\beta$
between 1.1 and 1.5. The filled and open dots are for systems
with spin parameter $\lambda < 0.08$ and $\lambda>0.08$, respectively.
The two solid lines roughly indicate the four magnitude scatter
in observations.
}
\end{figure}

For given set of $\alpha$ and $\beta$, 
the bulge-to-total ratio depends only on 
the spin parameter (cf. Fig. 1), 
our model therefore predicts a deterministic
relation (without any scatter) between
the bulge-to-total mass ratio and disk size 
(because $r_d \propto \lambda$, see MMW).
To see the scatter induced by the 
scatters in $\alpha$ and $\beta$, we plot,
in Fig. 4, the bulge-to-total mass ratio as a function of 
$r_d/L^{1/3}$, where $L$ is the total luminosity. 
According to the disk model we are considering here,
the quantity $r_d/L^{1/3}$ is expected to be approximately 
proportional to the spin parameter 
(see Mao, Mo \& White 1998 for a detailed discussion). 
The scatter plot is obtained from the same Monte Carlo simulation 
as shown in the bottom panel of Fig. 3. As one can see, 
there is a broad trend of $M_b/\Mtot$ with $r_d/L^{1/3}$ in the model
prediction, although the scatter in the relation is 
very large. For comparison, we plot (as the solid dots)
the observational data from the de Jong's sample. 
The model prediction is generally consistent with 
the observational result, but there are noticable differences.
There are more predicted systems at both the low and high ends 
of $M_b/\Mtot$, for the reasons we discussed before. There are
also several observed galaxies which have large 
$M_b/\Mtot$ and large $r_d/L^{1/3}$. Some of these galaxies 
seem to have low-surface-brightness disks. Their disks  
may appear fainter for their mass (and show up at the upper right part
of the diagram) because they have systematically higher mass-to-light
ratio (McGaugh \& de Block 1997).
Alternatively, these systems could form via other mechanisms
such as galaxy merging or their halos have extreme density and angular
momentum profiles. Clearly to make more detailed comparison, it is
important to obtain more observational data and 
also to understand the real
distributions of $\alpha$ and $\beta$ from future numerical
simulations.

\begin{figure}
\epsfysize=9.5cm
\centerline{\epsfbox{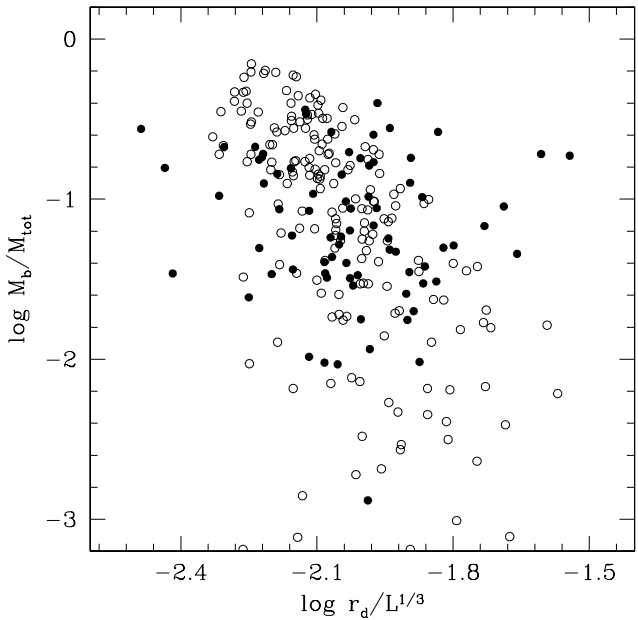}}
\caption{Monte Carlo results for the bulge-to-total mass ratio vs.
$r_d/L^{1/3}$, which is an approximate measure for the
spin parameter. We have allowed $\alpha$ to vary uniformly between 
0.7 to 1.3 and $\beta$ between 1.1 and 1.5.
The open circles are for the Monte Carlo results while the filled
dots are for the de Jong sample.
}
\end{figure}

\section {Discussion}

We have examined a scenario in which
the bulge-to-disc ratios of spiral galaxies 
are primarily determined by the angular momenta of 
their host dark haloes in the framework of
hierarchical structure formation models. The low angular momentum
gas in a dark halo falls into the centre and,  if it becomes self-gravitating
before settling into a rotationally supported disc, forms a bulge. 
The predicted bulge-to-disc ratios in both
size and luminosity are consistent with observational results.
In this model, the bulge and disc components in a galaxy 
have similar properties, because both form in a similar 
manner. This is consistent with the observations
of Courteau et al (1996) who showed that the colours and
other photometric properties of the bulge and disc components
are similar. This is also consistent with the fact that 
many galactic bulges are supported (at least partially) by
rotation rather than by random motion (Kormendy \& Illingworth, 1982;
Wyse et al 1997 and references therein).
Our model also predicts that low surface brightness galaxies have
systematically small bulges, in agreement with the observations
of O'Neil et al (1997) that many such galaxies 
lack bright central bulges.
Since the $\lambda$ distribution has only a weak dependence 
on halo mass (Lemson \& Kauffmann 1998), 
our model does not predict a correlation between 
the bulge-to-disc ratios and galaxy luminosities.
However, there is some weak observational evidence that
brighter galaxies have systematically larger
bulge-to-disc ratios (Schechter \& Dressler 1987).
Such a correlation can be produced by environmental effects: 
systems with higher mass may be biased towards higher density 
regions where discs can be truncated by tidal interaction
or gas stripping. Another possibility is that bulges are
easier to form in higher mass systems because their dark haloes
are less concentrated. 

We emphasize that the scenario explored here  
is unlikely to be the only channel for bulge
formation. All the processes discussed in the introduction 
may be relevant to some degree for the formation of bulges.
In particular, there is evidence that merging must have played 
some role in the formation of some massive bulges. 
The question is, of course, which processes are relevant
for the formation of the main population of bulges.
The merits of our model are: (1) it is built in the generally
successful framework of hierarchical clustering,
so that our assumptions can be tested by future 
numerical simulations; (2) it makes quantitative
predictions for the properties of the bulge-disc systems,
so that it can be falsified by future observations. 

\section*{Acknowledgments}
We thank Roelef de Jong for making his data available to us in
electronic form.
We are grateful to Simon White for helpful comments on the paper.
This project is partly supported by
the ``Sonderforschungsbereich 375-95 f\"ur Astro-Teilchenphysik'' der
Deutschen Forschungsgemeinschaft. 

{}
\end{document}